\documentclass[10pt,a4paper, twoside]{revtex4}
\usepackage{latexsym}
\usepackage[english]{babel}
\usepackage{amssymb}
\usepackage{amsfonts}
\usepackage{amsmath}
\usepackage{amsthm}
\usepackage{amscd}

\DeclareFixedFont{\sfracFont}{U}{euf}{b}{n}{7pt}
\newtheoremstyle{mydefi}% name
  {15pt}%      Space above
  {15pt}%      Space below
  {}%         Body font
  {}%         Indent amount (empty = no indent, \parindent = para indent)
  {\bfseries}% Thm head font
  {:}%        Punctuation after thm head
  {.5em}%     Space after thm head: " " = normal interword space;
  {}%         Thm head spec (can be left empty, meaning `normal')
\newtheoremstyle{mytheo}% name
  {15pt}%      Space above
  {15pt}%      Space below
  {\slshape}%         Body font
  {}%         Indent amount (empty = no indent, \parindent = para indent)
  {\bfseries}% Thm head font
  {:}%        Punctuation after thm head
  {.5em}%     Space after thm head: " " = normal interword space;
  {}%         Thm head spec (can be left empty, meaning `normal')

\theoremstyle{mytheo}
\newtheorem{stel}{Theorem}[section]
\newtheorem{lem}[stel]{Lemma}

\theoremstyle{mydefi}
\newtheorem{de}[stel]{Definition}

\newcommand{\BB}[1]{\mathbb{#1}}

\newcommand{\p}{|\!|}
\newcommand{\ten}{\otimes}
\newcommand{\EC}{\mathcal{E}}
\newcommand{\JC}{\mathcal{J}}
\newcommand{\MC}{\mathcal{M}}
\newcommand{\WC}{\mathcal{W}}
\newcommand{\Id}{\mbox{Id}}
\newcommand{\Ad}{\mbox{Ad}}

\newdimen\mylistindent \mylistindent=20pt  % inspringafstand van \myitem
\newcommand{\myitem}[1]{\par\noindent\everypar{\hangindent0cm\everypar{}}%
	\hangindent\mylistindent
	\hskip\mylistindent
	\relax\llap{#1\enspace}\ignorespaces}

\begin{document}
\title{\large{Constructing the Davies process of Resonance
Fluorescence with Quantum Stochastic Calculus}}
\author{\large{Luc Bouten$^\dagger$, Hans Maassen$^\dagger$ and Burkhard K\"ummerer$^{\dagger\dagger}$}} 
\affiliation{\large{
$^\dagger$Department of Mathematics, University of Nijmegen, Toernooiveld 1, 6525 ED Nijmegen, The Netherlands.\\
$^{\dagger\dagger}$Fachbereich Mathematik,
Technische Universit\"at Darmstadt,
Schlo\ss gartenstra\ss e 7,
64289 Darmstadt,
Germany.}}

\begin{abstract}
Starting point is a given semigroup of completely
positive maps on the $2\times 2$ matrices. This
semigroup describes the irreversible evolution of
a decaying $2$-level atom.
Using the integral-sum kernel approach \cite{Maa}
to quantum stochastic calculus we couple the
$2$-level atom to an environment, which in our
case will be interpreted as the electromagnetic field.
The irreversible time evolution of the $2$-level atom
then stems from the reversible time evolution of atom 
and field together. Mathematically speaking, we have
constructed a Markov dilation \cite{Kum1} of the semigroup.

The next step is to drive the atom by a laser and
to count the photons emitted into the field by the 
decaying $2$-level atom. For every possible sequence of 
photon counts we construct a map that gives the time 
evolution of the $2$-level atom inferred by that sequence.
The family of maps that we obtain 
in this way forms a so-called Davies process \cite{Dav}, \cite{SrD}.
In his book Davies describes the structure 
of these processes, which brings us into the field 
of quantum trajectories \cite{Car}. Within our model 
we calculate the jump operators and we briefly 
describe the resulting counting process.
\end{abstract}

\maketitle

\large{
\section{Introduction}

In this paper we want to illustrate that quantum stochastic calculus together
with the processes studied by Davies in his book \cite{Dav}, and explained in his paper with 
Srinivas \cite{SrD}, form a suitable mathematically rigorous framework for doing 
quantum trajectory theory \cite{Car}. As an example we consider here the case of 
resonance fluorescence.                                                                         \\
Our starting point is a semigroup of transition operators $\{T_t\}_{t\ge 0}$ on
the algebra $M_2$ of all $2\times 2$-matrices. This semigroup describes the irreversible
evolution of a spontaneously decaying $2$-level atom in the Heisenberg picture. By
coupling the atom to a quantum noise, we construct a stationary 
quantum Markov process having precisely these transition operators. If we impose the
requirements that the external noise be a Bose field, and the quantum Markov process be
minimal, then the latter is uniquely determined. It is called the \emph{minimal Bose dilation}
of $(M_2, T_t, g)$ \cite{Kum2}, where $g$ is the ground state of the $2$-level atom.                     \\
Since this dilation is uniquely determined, any other reversible dynamical model which
couples $(M_2, T_t, g)$ to some Bose field necessarily contains this Bose dilation as a
subsystem. Therefore, without deriving our model from an explicit Schr\"odinger equation
(by performing a Markovian limit) we may safely assume it to be a physically correct way
to describe the interaction of the $2$-level atom with the electromagnetic field.               \\
We will couple the $2$-level atom to the electromagnetic field by using quantum stochastic
calculus \cite{Par}, \cite{Mey}. We use a version of quantum stochastic calculus based on
integral-sum kernels \cite{Maa}, \cite{LiM}, \cite{Mey}, which has the advantage that we
have an explicit construction for the solution of the quantum stochastic differential
equation with which we will describe the coupling of atom and field. 
Having this explicit construction in our hands is important for doing the actual calculations
we encounter later on.                                                                           \\
To be able to discuss resonance fluorescence we have to use a dilation where we have two
channels in the electromagnetic field. On one of them we will put a laser state to drive the 
$2$-level atom. We will call this field the \emph{forward channel} and the other one the \emph{side channel}.
We will then count photons in both channels. We need the side channel, because there we know
that all detected photons are fluorescence photons. In the forward channel a detected photon
could just as well be coming directly from the laser.                                              \\
For every event that can occur in the photon counters we construct a map giving the
evolution of the $2$-level atom inferred by that event. We will see that the family of maps we
obtain, fulfills the axioms for the processes discussed by Davies \cite{Dav}. We have constructed
the Davies process of resonance fluorescence.                                                       \\
Using the structure theory for Davies processes \cite{Dav} we can decompose the process into
its trajectories \cite{Car}. Within our model we calculate the expression for the
jump operators and for the time evolution in between jumps. Note that a jump in the system
occurs the moment we detect a photon, since our knowledge concerning the system changes.              \\
Using the above apparatus we show that the resulting counting process in the 
side channel is a so-called \emph{renewal process}.

\section{The dilation}

Let $M_2$, the algebra of $2 \times 2$-matrices, stand for the algebra of observables of
a $2$-level atom. On this algebra we are given a (continuous) semigroup $\{T_t\}_{t\ge 0}$
of completely positive maps. This semigroup describes the, generally irreversible,
evolution of the $2$-level atom. Lindblad's Theorem \cite{Lin} then says that $T_t = \exp{tL}$
where $L: M_2 \to M_2$ can be written as: for $A\in M_2$:
  \begin{equation}
  L(A) = i[H,A] + \sum_{j=1}^k V_j^*AV_j -\frac{1}{2}\{V_j^*V_j, A\}, 
  \end{equation}
where the $V_j$ and $H$ are fixed $2\times2$-matrices, $H$ being Hermitian. In this paper we will 
restrict to the simpler case where $H = 0$ and there are just two $V_j's$. This means there
is dissipation only into two channels, the forward channel described by $V_f$, and the side channel
described by $V_s$. We choose $V_f$ and $V_s$ such that: 
  \begin{equation*}
  V= \begin{pmatrix} 0 & 0  \\ 1 & 0 \end{pmatrix},\mbox{\ \ \ } V_f = \kappa_f V,
  \mbox{\ \ \ }V_s = \kappa_s V, \mbox{\ \ \ } |\kappa_f|^2 + |\kappa_s|^2 = 1.
  \end{equation*}
This exactly gives the time evolution for spontaneous decay to the ground state of the $2$-level atom
into two decay channels, where the decay rates are given by $|\kappa_f|^2$ and $|\kappa_s|^2$.        \\
We want to see this irreversible evolution of the $2$-level atom as stemming from a reversible evolution
of the atom coupled to, in this case, two decay channels in the field.
So let us first construct the algebra of observables for these fields. 
Let $\mathcal{F}$ be the symmetric Fock space over the Hilbert space $L^2(\BB{R})$ of square
integrable wave functions on the real line, i.e.\ $\mathcal{F} := 
\BB{C}\oplus \bigoplus_{n=1}^\infty L^2(\BB{R})^{\ten_s n}$.
The electromagnetic field is given by creation and annihilation operators
on $\mathcal{F}$, generating the algebra of all bounded operators. We need two copies of this 
algebra, which we denote by $\WC_f$, the field which will be the forward channel, and $\WC_s$, the 
field which will be the side channel.                                                                 \\
The evolution over a time $t$ of a free field is given by the second quantization of the
left shift, i.e.\ the second quantization of the operator on $L^2(\BB{R})$ which maps $f(\cdot)$ into
$f(\cdot + t)$. We denote the second quantization of this operator by $S_t$. This means that in
the Heisenberg picture we have an evolution on $\WC_f \ten \WC_s$ mapping $A$ into
$(S_{t}^*\ten S_{t}^*)A(S_{t}\ten S_{t})$ \big(= $(S_{-t}\ten S_{-t})A(S_t\ten S_t)$\big),
also denoted by $\Ad[S_t\ten S_t](A)$.                                                                  \\
The presence of the atom in the fields introduces a perturbation on the evolution of the free
fields. We let this perturbation be given by a certain family of unitary operators
$\{U_t\}_{t\in \BB{R}}$ on $\BB{C}^2\ten\mathcal{F}\ten\mathcal{F}$, which will be specified later,
that forms a \emph{cocycle} with respect to the shift $S_t\ten S_t$, i.e.\ for all $t,s \in \BB{R}:\
U_{t+s} = (S_{-s}\ten S_{-s})U_t(S_{s}\ten S_{s})U_s$. Given this cocycle, we let the time evolution of
the atom and the fields together be given by the following one-parameter group $\{\hat{T}_t\}_{t\in\BB{R}}$
(i.e.\ the evolution is now \emph{reversible}) of $*$-automorphisms on $M_2 \ten \WC_f \ten \WC_s$:
for all $A \in M_2 \ten \WC_f \ten \WC_s$:
  \begin{equation*}
  \hat{T}_t(A) = \left\{ \begin{array}{ll}
  U_t^{-1}S_{-t}AS_tU_t & \mbox{\ \ \ if \ } t \ge 0 \\
  S_{-t}U_{-t}A U^{-1}_{-t}S_t & \mbox{\ \ \ if \ } t < 0 
  \end{array}\right. ,
  \end{equation*}                                                                                      \\
The solution of the following quantum stochastic differential equation 
\cite{HuP}, \cite{Par} provides us with a cocycle of unitaries with respect to the shift:
  \begin{equation}\label{HuPeq}
  dU_t = \{V_f dA^*_{f,t} -V_f^* dA_{f,t} + V_s dA^*_{s,t} - V_s^* dA_{s,t} -\frac{1}{2}V^*Vdt\}U_t,
  \mbox{\ \ \ } U_0 = I.
  \end{equation}
In the next section we will give an explicit construction for the solution $U_t$ of this equation.
It can be shown (\cite{HuP}, \cite{Fri}, \cite{Maa}, \cite{Par}) that if the cocycle satisfies
equation \eqref{HuPeq} we have constructed a so-called \emph{quantum Markov dilation}
$(M_2\ten\WC_f\ten \WC_s, \{\hat{T}\}_{t\in \BB{R}}, \mbox{id}\ten\phi\ten\phi)$ of the quantum
dynamical system $(M_2, \{T_t\}_{t \ge 0}, g)$ \cite{Kum1}, \cite{Kum2}, where $\phi$ is the
vector state on $\WC_{f,s}$ given by the vacuum vector. This means that the following
\emph{dilation diagram} commutes for all $t \ge 0$ (and that the resulting quantum process is Markov):
    \begin{equation}\label{dildiag}\begin{CD}
  M_2 @>T_t>> M_2              \\
   @V{\Id \ten I \ten I}VV        @AA{\Id \ten \phi \ten \phi}A      \\
   M_2\ten \WC_f \ten \WC_s @>\hat{T}_t>> M_2\ten\WC_f \ten\WC_s            \\
  \end{CD}\end{equation}
i.e.\ for all $A \in M_2:\ T_t(A) = \big(\Id \ten \phi \ten \phi\big) 
\big(\hat{T}_t(A \ten I \ten I)\big)$.  \\
Let us here look briefly in the Schr\"odinger picture at the above diagram. 
If we start with a state $\rho$ of the $2$-level atom (i.e.\ we are now in the upper right hand
corner of the diagram), then this state undergoes the following sequence of maps:
  \begin{equation*} \begin{split}
  &\rho(\cdot) \mapsto \rho\big( \Id \ten \phi \ten \phi(\cdot)\big) = 
  \rho \ten \phi \ten \phi(\cdot) \mapsto 
  \rho \ten \phi \ten \phi \big( \hat{T}_t (\cdot)\big)\mapsto             \\
  &\rho \ten \phi \ten \phi \big(\hat{T}_t(\cdot\ten I \ten I)\big) = 
  \mbox{Tr}_{\mathcal{F}_f \ten\mathcal{F}_s}\big(\hat{T}_{t*}(\rho \ten \phi \ten \phi)\big)(\cdot),
  \end{split}\end{equation*}
i.e.\ $\rho$ maps to $\mbox{Tr}_{\mathcal{F}_f \ten\mathcal{F}_s}\big(\hat{T}_{t*}(\rho \ten \phi 
\ten \phi)\big)$. This means that $\rho$ is first coupled to the two fields both in the vacuum state, 
then they are time evolved together and then there is a partial trace taken over the two fields.

\section{Guichardet space and integral-sum kernels}

Let us now turn to giving the explicit construction for the solution of equation \eqref{HuPeq}. 
For this we need the \emph{Guichardet space} $\Omega$ \cite{Gui} of $\BB{R}$, which is
the space of all finite subsets of $\BB{R}$, i.e.\ 
$\Omega := \bigcup_{n\in \BB{N}} \Omega_n$, where $\Omega_n := \{\sigma \subset \BB{R};\ |\sigma| = n \}$.
Let us denote by $\lambda_n$ the Lebesgue measure on $\BB{R}^n$.
If, for $n \in \BB{N}$, we let $j_n: \BB{R}^n \to \Omega_n$ denote the map that maps an $n$-tuple
$(t_1,t_2,\ldots,t_n)$ into the set $\{t_1,t_2, \ldots, t_n\}$, then we can define a measure $\mu_n$
on $\Omega$ by: $\mu_n(E) := \frac{1}{n!}\lambda_n\big(j_n^{-1}(E)\big)$ for all $E$ in the sigma field
$\Sigma_n$ of $\Omega_n$ induced by $j_n$ and the Borel sigma field of $\BB{R}^n$. Now we define a measure
$\mu$ on $\Omega$ such that $\mu(\{\emptyset\}) = 1$ and $\mu = \mu_n$ on $\Omega_n$. This means we have
now turned the Guichardet space into the measure space $(\Omega, \Sigma, \mu)$.                          \\
The key to constructing the solution of equation \eqref{HuPeq} is to identify the symmetric Fock space
$\mathcal{F}$ with the space of all quadratically integrable functions on the Guichardet space
$L^2(\Omega, \mu)$. To see this identification note that $L^2(\Omega_n, \mu_n)$ is, in the canonical
way, unitarily equivalent with the space of all quadratically integrable functions on $\BB{R}^n$
invariant under permutations of coordinates, denoted $L_{\mbox{sym}}^2(\BB{R}^n)$. It is now
obvious how to identify $\mathcal{F} = \BB{C} \oplus \bigoplus_{n = 1}^\infty L^2_{\mbox{sym}}(\BB{R}^n)$
with $L^2(\Omega, \mu) = \BB{C} \oplus \bigoplus_{n=1}^\infty L^2(\Omega_n, \mu_n)$.                        \\
For every $f \in L^2(\BB{R})$ we define the \emph{exponential vector} $e(f) \in \mathcal{F}$ in the
following way: $e(f) := 1 \oplus f \oplus \frac{1}{\sqrt{2}} f^{\ten 2} \oplus 
\frac{1}{\sqrt{6}} f^{\ten 3} \oplus \dots$. Note that the linear span of all exponential vectors
forms a dense subspace of $\mathcal{F}$. For every $f \in L^2(\BB{R})$ we define the \emph{coherent
vector} $\psi(f)$ to be the exponential vector of $f$ normalised to unity, i.e.\ $\psi(f) =
\exp(-\frac{1}{2}\p f\p^2)e(f)$. Under the above identification of $\mathcal{F}$ with
$L^2(\Omega, \mu)$, the exponential vector (of an $f \in L^2(\BB{R})$) $e(f)$ is mapped into an element
of $L^2(\Omega, \mu)$ which we denote by $\pi(f)$ and which is given by: $\pi(f):\ \Omega \to \BB{C}:\
\omega \mapsto \prod_{s \in \omega} f(s)$, where the empty product $\prod_{s \in \emptyset}f(s)$ is
defined to be $1$. We will often choose for $f$ the \emph{indicator function} of a certain interval
$I \subset \BB{R}$, which we denote by $\chi_I$. This is the function which is $1$ on $I$ and
$0$ elsewhere.                                                                                        \\
We are now ready to start the construction of the solution $U_t$ of equation $\eqref{HuPeq}$. Define
the \emph{integral-sum kernel} of $U_t$ (name will become apparent in a minute)
to be the map $u_t$ that maps four disjoint finite subsets of $\BB{R}, \sigma_f, \sigma_s, \tau_f, \tau_s$
(where $f$ and $s$ stand for "forward" and "side") to the following $2 \times 2$-matrix, where we write
$\sigma_f \cup \sigma_s\cup \tau_f \cup \tau_s$ also as $\{t_1, t_2, \ldots, t_k\}$ such that
$t_1 < t_2 < \ldots < t_k$ and $k \in \BB{N}$: 
  \begin{equation*}\begin{split}
  u_t(\sigma_f, \sigma_s, \tau_f, \tau_s) := &\pi(\chi_{[0,t]})(\sigma_f \cup \sigma_s\cup \tau_f \cup \tau_s)
  \exp(- \frac{t-t_k}{2}V^*V)V_k \times                                                      \\
  &\exp(- \frac{t_k -t_{k-1}}{2}V^*V)V_{k-1}\ldots V_1 \exp(-\frac{t_1}{2}V^*V),
  \end{split}\end{equation*}
where
  \begin{equation*}
  V_j = \left\{ \begin{array}{ll}
  V_f & \mbox{\ \ if \ } t_j \in \sigma_f \\
  -V_f^* & \mbox{\ \  if \ } t_j \in \tau_f \\
  V_s & \mbox{\ \ if \ } t_j \in \sigma_s \\
  -V_s^* & \mbox{\ \  if \ } t_j \in \tau_s
  \end{array}\right. .
  \end{equation*}
Then we have the following theorem of Maassen, see \cite{Maa}, \cite{LiM}:
\begin{stel}\label{Maassen}
After identifying $\BB{C}^2 \ten \mathcal{F}\ten\mathcal{F}$ with $L^2_{\BB{C}^2}(\Omega\times\Omega, \mu\times\mu)$,
the space of all square integrable functions on $\Omega \times \Omega$ with values in $\BB{C}^2$,
the solution $U_t:\ L^2_{\BB{C}^2}(\Omega\times\Omega, \mu\times\mu) \to L^2_{\BB{C}^2}(\Omega\times\Omega, \mu\times\mu)$
of equation \eqref{HuPeq} is given by:
  \begin{equation*}
  (U_tf)(\omega_f, \omega_s) = \sum_{\substack{\sigma_f\subset\omega_f\\ \sigma_s\subset\omega_s}}
  \int_{\Omega\times\Omega} u_t(\sigma_f,\sigma_s,\tau_f,\tau_s)               
  f\big((\omega_f\backslash\sigma_f)\cup\tau_f, (\omega_s\backslash\sigma_s)\cup\tau_s\big)
  d\tau_f d\tau_s.
  \end{equation*} 
\end{stel}

Now we have an explicit expression for the time evolution $\hat{T}_t = \Ad[\hat{U}_t]$, where $\hat{U}_t$
is given by $S_tU_t$ if $t\ge 0$ and $U_{-t}^{-1}S_{t}$ if $t <0$. The family $\{\hat{U}_t\}_{t\in \BB{R}}$
forms a group of unitary operators on $\BB{C}^2 \ten \mathcal{F}\ten\mathcal{F}$ describing the time
evolution of the $2$-level atom and the two fields together. Stone's Theorem says that there must be
a Hamiltonian associated to this time evolution. This Hamiltonian has been calculated recently \cite{Gre}.

\section{The Davies process}

We now return to the situation in figure \ref{dildiag}. We wish to make some changes in this diagram
and for this we need to introduce some more notation regarding Guichardet spaces. Let $I \subset \BB{R}$
be an interval. Then the \emph{Guichardet space of $I$} is the set $\Omega(I) =
\bigcup_{n=0}^\infty \Omega_n(I)$, where $\Omega_n(I) = \{\sigma \subset I;\ |\sigma| = n\}$. In
a similar way as for $\Omega$, which is $\Omega(\BB{R})$, we can give these sets a measure structure:
$(\Omega(I), \Sigma(I), \mu)$. Given a subset $E$ of $\Omega(I)$ in the sigma field $\Sigma(I)$, we
can construct the projection $M_{\chi_E}:\ L^2(\Omega, \mu) \to L^2(\Omega, \mu):\ f \mapsto \chi_E f$.  \\
Let $I$ be $[-t,0)$, then the events in $\Sigma\big([-t,0)\big)$, which we abbreviate to $\Sigma_t$,
are events in the output field of the atom up to time $t$. Remember that the evolution of a free
field was given by the left shift and that the atom is sitting in the origin. 
Since the Guichardet space representation corresponds to the photon number picture, we can give concrete
interpretations to the subsets in $\Sigma_t$. For instance, the subsets $\Omega_n\big([-t,0)\big)$,
correspond to the events "there are n photons in the output of the atom into this field up to time t".        \\
Now back to the situation in figure \ref{dildiag}. Suppose we have been observing the output in the
forward and side channel of the atom up to time $t$ with two photon counters. Then we are given two
events $E_f$ and $E_s$ in $\Sigma_t$. Since we know the outcome of the measurements we
have to change the time evolution of the $2$-level atom, i.e.\ we have to project onto the observed events
(see also \cite{BaB}).
This is summarized in the following figure:       
  \begin{equation*}\begin{CD}
  M_2 @>\EC^t_0[E_f, E_s]>> M_2              \\
   @V{\Id \ten \chi_{E_f} \ten \chi_{E_s}}VV        @AA{\Id \ten \phi \ten \phi}A      \\
   M_2\ten \WC_f \ten \WC_s @>\hat{T}_t>> M_2\ten\WC_f \ten\WC_s            \\
  \end{CD}\end{equation*}
where we have suppressed the capital letters $M$ in the projections. The map $\EC_0^t[E_f, E_s]:\ M_2 \to M_2:\
A \mapsto  \Id \ten \phi \ten \phi\big(\hat{T}_t(A\ten\chi_{E_f} \ten \chi_{E_s})\big)$ is
the unnormalized time evolution of the $2$-level atom in the Heisenberg picture given that we see
event $E_f$ in the output of the forward channel and event $E_s$ in the output of the side channel.
If we are given a state on $M_2$, i.e.\ a $2 \times 2$ density matrix $\rho$, then the probability
of seeing event $E_f$ in the forward channel and $E_s$ in the side channel after $t$ seconds of
observation is given by: $\BB{P}^t_\rho[(E_f, E_s)] = \mbox{Tr}\big(\rho\EC_0^t[E_f, E_s](I)\big)$.     \\
The setting is still not complete for describing resonance fluorescence. Since we are not driving the
atom, both the forward and the side channel fields are in the vacuum state, at most one photon can appear 
in the output. We change this by putting on the forward channel a coherent state with
amplitude $z \in \BB{C}$, defined by: $\gamma_{z_t}:\ \WC \to \BB{C}:\ A \mapsto
\exp(-t|z|^2)\big\langle \pi(z\chi_{[0,t]}), A\pi(z\chi_{[0,t]})\big\rangle$. Note that $\gamma_0$ is
the vacuum state. Putting a coherent state on the forward channel mimics a laser driving the atom. 
We have suppressed its oscillations for the sake of simplicity. Now we are ready to do 
resonance fluorescence, i.e.\ the diagram has changed into:
  \begin{equation*}\begin{CD}
  M_2 @>\EC^t_z[E_f, E_s]>> M_2              \\
   @V{\Id \ten \chi_{E_f} \ten \chi_{E_s}}VV        @AA{\Id \ten \gamma_{z_t} 
\ten \gamma_0}A      \\
   M_2\ten \WC_f \ten \WC_s @>\hat{T}_t>> M_2\ten\WC_f \ten\WC_s            \\
  \end{CD}\end{equation*} 
where the map $\EC_z^t[E_f, E_s]:\ M_2 \to M_2$ is now defined by $\EC_z^t[E_f, E_s](A) :=
\Id \ten \gamma_{z_t} \ten \gamma_0 \big(\hat{T}_t(A\ten\chi_{E_f} \ten \chi_{E_s})\big)$. It describes
the unnormalized time evolution of the laser-driven atom given that we see event $E_f$ in the output of
the forward channel and event $E_s$ in the output of the side channel. Given a state $\rho$ of the atom,
the probability of seeing event $E_f$ in the forward channel and $E_s$ in the side channel after $t$ seconds of
observation is now given by: $\BB{P}^t_\rho[(E_f, E_s)] = \mbox{Tr}\big(\rho\EC_z^t[E_f, E_s](I)\big)$.
To make the notation lighter we suppres the $z$ in $\EC^t_z$ in the following.                        \\
Since $L^2(\Omega, \Sigma, \mu) \ten L^2(\Omega, \Sigma, \mu)$ is canonically isomorphic to
$L^2 (\Omega\times\Omega, \Sigma \ten \Sigma, \mu \times\mu)$ we can simplify our notation even a bit
further. By identifying these spaces we can write $\EC^t[E_f, E_s] = \EC^t[E_f\times E_s]$, where
the righthandside is defined by: for all $E \in \Sigma_t \ten \Sigma_t, A \in M_2, t\ge 0:\
\EC^t[E](A) := \Id \ten \gamma_{z_t,0} \big(\hat{T}_t(A\ten\chi_E)\big)$, where $\gamma_{z_t,0}$
is an abbreviation for $\gamma_{z_t}\ten \gamma_0$. We will now study the properties of the family 
of maps we defined.

\begin{stel}\label{properties} 
The family of maps $\{\EC^t[E]\}_{t\ge0, E \in \Sigma_t \ten \Sigma_t}$ satisfies
the axioms of a \emph{Davies process}, \cite{Dav}: 
\setlength{\parskip}{0.5\baselineskip}
  \myitem{1.}  For all $t \ge 0$ and $E \in \Sigma_t \ten \Sigma_t$, $\EC^t[E]$ is completely positive.       
  \myitem{2.}  For all $t \ge 0$ and all countable collections of disjoint sets 
               $\{E_n\}$ in $\Sigma_t \ten \Sigma_t$ \\ 
               and for all $A \in M_2:\
               \EC^t\Big[\bigcup_{n}E_n\Big](A) = \sum_{n} \EC^t[E_n](A)$.
  \myitem{3.}  For all $t \ge 0$ we have 
               $\EC^t\Big[\Omega\big([-t,0)\big)\times\Omega\big([-t,0)\big)\Big] (I) = I$.
  \myitem{4.}  For all $A \in M_2:\ \lim_{t \to 0}
               \EC^t\Big[\Omega\big([-t,0)\big)\times\Omega\big([-t,0)\big)\Big] (A) = A$.
  \myitem{5.}  For all $t,s \ge 0$ and $E \in \Sigma_s \ten \Sigma_s, 
               F \in \Sigma_t \ten \Sigma_t$ and all $A \in M_2$ we have:\\
               $\EC^t[F]\circ\EC^s[E](A) = \EC^{s+t}[F-s \tilde{\cup} E] (A)$, \\
               where $F-s \in \Sigma\big([-t-s, -s) \ten \Sigma\big([-t-s, -s) \big)$
               is given by:                                                              \\
               $F-s = \{(f_f-s,f_s-s);\ (f_f,f_s) \in F)\}$ and $\tilde{\cup}$
               is defined by:                                                    \\
               $A \tilde{\cup} B = \{(\omega_f \cup \sigma_f,\omega_s \cup \sigma_s);\ 
               (\omega_f,\omega_s)\in A, (\sigma_f,\sigma_s) \in B\}$.
\setlength{\parskip}{\baselineskip}
\end{stel}

\begin{proof}
The only point where there is really something to prove is point $5$.
Let us first introduce some short notation which we shall only use in this proof.
Let $\pi(z_t, 0)$ denote $\pi(z\chi_{[0,t]}) \ten \pi(0)$ and denote
$S_t \ten S_t$ just by $S_t$. Further we use the notation $\sigma_t(U_s)$ for $S_{-t}U_sS_t$.
Then for all $A \in M_2$, $s,t \ge 0$, $E\in \Sigma_s\ten\Sigma_s$ and $F \in \Sigma_t\ten \Sigma_t$ we have:
  \begin{equation*}\begin{split}
  &\frac{\EC^t[F]\circ\EC^s[E](A)}{\exp(-(s+t)|z|^2)} = 
     \EC^t[F]\Big(\big\langle\pi(z_s,0),\hat{T}_s(A\ten\chi_E)    
      \pi(z_s,0)\big\rangle\Big)\exp(t|z|^2) =                                      \\
  &\Big\langle\pi(z_t,0), \hat{T}_t\Big(\big\langle\pi(z_s,0),\hat{T}_s(A\ten\chi_E)
      \pi(z_s,0)\big\rangle\ten\chi_F\Big)\pi(z_t,0)\Big\rangle =                            \\
  &\Big\langle\pi(z_t,0), U_t^*\big\langle\pi(z_s,0),\hat{T}_s(A\ten\chi_E)    
      \pi(z_s,0)\big\rangle\ten\chi_{F+t} U_t\pi(z_t,0)\Big\rangle =                            \\
  &\Big\langle\pi(z_t,0), U_t^*\big\langle S_{-t}\pi(z_s,0),S_{-t}\hat{T}_s(A\ten\chi_E)
      S_t S_{-t}\pi(z_s,0)\big\rangle\ten\chi_{F+t} U_t\pi(z_t,0)\Big\rangle =                            \\
  &\Big\langle\pi(z_t,0), U_t^*\big\langle S_{-t}\pi(z_s,0),\sigma_t(U_s)^*A\ten\chi_{E+t+s}
      \sigma_t(U_s) S_{-t}\pi(z_s,0)\big\rangle\ten\chi_{F+t} U_t\pi(z_t,0)\Big\rangle.
  \end{split}\end{equation*} 
Now we use the cocycle identity and the continuous tensor product structure of the symmetric Fock
space to obtain:
  \begin{equation*}\begin{split}
  &\frac{\EC^t[F]\circ\EC^s[E](A)}{\exp(-(s+t)|z|^2)} = 
  \Big\langle \pi(z_{t+s},0), (\sigma_t(U_s)U_t)^* 
    A \ten \chi_{F+t \tilde{\cup} E+t+s}\sigma_t(U_s)U_t\pi(z_{t+s},0)\Big\rangle  = \\
  &\Big\langle \pi(z_{t+s},0), U_{t+s}^*
    A \ten \chi_{F+t \tilde{\cup} E+t+s}U_{t+s}\pi(z_{t+s},0)\Big\rangle = \\ 
  &\Big\langle \pi(z_{t+s},0), \hat{T}_{t+s}
    (A \ten \chi_{F-s \tilde{\cup} E})\pi(z_{t+s},0)\Big\rangle = 
  \frac{\EC^{s+t}[F-s \tilde{\cup} E](A)}{\exp(-(s+t)|z|^2)}.
  \end{split}\end{equation*}
\end{proof}

Define maps $Y_t: M_2 \to M_2: A \mapsto \EC^t\big[\{(\emptyset,\emptyset)\}\big](A)$. They represent 
the evolution of the atom when it is observed that no photons entered the decay channels. Then we have that 
for all $t,s \ge 0:\ Y_tY_s = \EC^t\big[\{(\emptyset,\emptyset)\}\big]\circ
\EC^s\big[\{(\emptyset, \emptyset)\}\big]= \EC^{t+s}\big[\{(\emptyset, 
\emptyset)\}-s \tilde{\cup}\{(\emptyset, \emptyset)\}\big] =
\EC^{t+s}\big[\{(\emptyset, \emptyset)\}\big] = Y_{t+s}$, i.e.\ the family $\{Y_t\}_{t\ge 0}$ forms
a semigroup. \\
Now observe that for $A \in M_2$ and $t \ge 0$ we have: 
  \begin{equation*}\begin{split}
  &Y_t(A) = \EC^t\big[(\{\emptyset\},\{\emptyset\})\big](A) = 
  \Id\ten \gamma_{z_t}\ten \gamma_0\big(\hat{T}_t(A\ten\chi_{\{\emptyset\}}\ten\chi_{\{\emptyset\}})\big) = \\
  &\big\langle \pi(z_t)\ten\pi(0),U_t^*A\ten\chi_{\{\emptyset\}}\ten\chi_{\{\emptyset\}}
  U_t\pi(z_t)\ten\pi(0)\big\rangle\exp(-t|z|^2) =                                                            \\
  &\big(U_t\pi(z_t)\ten\pi(0)\big)^*(\emptyset,\emptyset)A\big(U_t\pi(z_t)\ten\pi(0)\big)
   (\emptyset,\emptyset)\exp(-t|z|^2).
  \end{split}\end{equation*}
If we define $B_t: \BB{C}^2 \to \BB{C}^2:\ v 
\mapsto \Big(\exp(-\frac{1}{2}t|z|^2)\big(U_t\pi(z_t)\ten\pi(0)\big)\Big)
(\emptyset,\emptyset)v$, then we see, using Theorem \ref{Maassen}, that $B_t$ is the following semigroup
of contractions:
  \begin{equation}\label{Bt}
  B_t = \exp\big(-\frac{1}{2}(|z|^2I_2 + V^*V +2zV_f^*)t\big),
  \end{equation}
and for all $A \in M_2:\ Y_t(A) = B_t^*AB_t$. We say that the Davies process $\EC^t$ is \emph{ideal},
see \cite{Dav}.                                                                                        \\
Furthermore Theorem \ref{properties} point $2$, leads to:
  \begin{equation*}
  \EC^t\big[\Omega[-t,0)\times \Omega[-t,0)\backslash \{(\emptyset, \emptyset)\}\big] (I) = I - B_t^*B_t.
  \end{equation*}
If we use this and the expression for $B_t$ \eqref{Bt}, then we can do some estimations which
in the end lead to:
  \begin{equation*}
  \EC^t\big[\Omega[-t,0)\times \Omega[-t,0)\backslash \{(\emptyset, \emptyset)\}\big] (I) \le tKI,
  \end{equation*}
with $K = (2|z|^2|\kappa_f|^2 +1)$. This property can be summarized by saying that the Davies process
$\EC^t$ has \emph{bounded interaction rate}, see \cite{Dav}.

\section{Quantum trajectories}

In the seventies Davies studied the structure of what we now call Davies processes \cite{Dav}.
Let us first state his results, as far as relevant, in the context of the process we are studying.
\begin{lem}\label{lemma}\textbf{(Davies \cite{Dav})}
Given any ideal Davies process $\EC^t$ with bounded interaction rate, as defined in the previous
section, we have existence of the following limits:
  \begin{equation*}
  \JC_f := \lim_{t \downarrow 0} \frac{1}{t}\EC^t\big[\Omega_1[-t,0), \{\emptyset\}\big] \mbox{\ \ and \ \ }
  \JC_s := \lim_{t \downarrow 0} \frac{1}{t}\EC^t\big[\{\emptyset\},\Omega_1[-t,0)\big].
  \end{equation*}
\end{lem}  
These completely positive maps represent the action we have to apply on the $2$-level atom the moment
we see one photon appear in the forward, respectively side channel.They are the \emph{jump operations} for
these channels. We will explicitly calculate these limits later on, but first we turn our attention to
decomposing the Davies process into its trajectories \cite{Car}. For this we need the following
definition. 
\begin{de}\label{W}
Let $Y_t: M_2 \to M_2$ be the maps from the previous section, i.e. $Y_t = \EC^t[\{\emptyset\},\{\emptyset\}]$
and let $\JC_f$ and $\JC_s$ be the maps from lemma \ref{lemma}. Let $\omega_f$ and $\omega_s$ be disjoint
elements of $\Omega[-t,0)$ and denote $\omega_f \cup \omega_s$ also as $\{t_1, \ldots, t_k\}$ where
$-t \le t_1 < t_2 < \ldots < t_k \le 0$ for a $k \in \BB{N}$. Then we define:
  \begin{equation*}
  W_{Y,\JC_f, \JC_s}(\omega_f, \omega_s)
         := Y_{t_1+t}\JC^{t_1}Y_{t_2-t_1}\JC^{t_2}\ldots \JC^{t_k}Y_{-t_k}, 
  \end{equation*}
where $\JC^{t_i}$ denotes $\JC_s$ if $t_i \in \omega_s$ and $\JC_f$ if $t_i \in \omega_f$.
\end{de}
Since $Y_t$ is the time evolution of the system when, both in the forward and the side channels,
no photons are detected and $\JC_f$ and $\JC_s$ are the jump operations that we have to apply when a 
photon in the corresponding channels appears, it is clear that the string of maps 
$Y_{t-t_1}\JC^{t_1}Y_{t_1-t_2}\JC^{t_2}\ldots \JC^{t_k}Y_{t_k}$ represents the
\emph{trajectory} of an observable $A$ in $M_2$ when we find the outcomes $\omega_f$ in the
forward and $\omega_s$ in the side channel during our counting experiment. The following theorem
of Davies \cite{Dav} shows how to decompose the Davies process into its trajectories.
\begin{stel}\textbf{(Davies \cite{Dav})}\label{deco} 
Given any ideal Davies process $\EC^t$ with bounded interaction rate, as defined in the previous
section, we have for all $t \ge 0$, $E_f, E_s \in \Sigma_t$ and $A \in M_2$:
  \begin{equation*}
  \EC^t[E_f,E_s](A) = \int_{E_f \times E_s} W_{Y,\JC_f, \JC_s}(\omega_f, \omega_s) (A)d\mu(\omega_f)d\mu(\omega_s).
  \end{equation*}
\end{stel}
In the previous section we already found the expression for the time evolution in between jumps: $Y_t$. We
now turn to the calculation of $\JC_f$ and $\JC_s$. For all $A$ in $M_2$ we have:
  \begin{equation*}
  \JC_f(A) = \lim_{t\downarrow 0} \frac{1}{t} \EC^t\big[(\Omega_1[-t,0),\{\emptyset\})\big](A)
  = \lim_{t\downarrow 0} \frac{\int_0^t\Ad\big[U_t\pi(z)\ten\pi(0)(\{s\},\emptyset)\big](A)ds}
       {t\exp(-t|z|^2)}.                                  
  \end{equation*}
Now look at $U_t\pi(z)\ten\pi(0)(\{s\},\emptyset)$, we find by using Theorem \ref{Maassen}:
  \begin{equation*}\begin{split}
  &U_t\pi(z)\ten\pi(0)(\{s\},\emptyset) =                                                  
  \sum_{\sigma\subset\{s\}}\int_{\Omega}u_t(\sigma,\emptyset,\tau,\emptyset)
  z^{1-|\sigma|+|\tau|}d\tau = zu_t(\emptyset,\emptyset,\emptyset,\emptyset)\ +        \\
  &z^2\int_0^tu_t(\emptyset,\emptyset,\{r\},\emptyset)dr\ +                    
  u_t(\{s\},\emptyset,\emptyset,\emptyset) + z\int_0^tu_t(\{s\},\emptyset,\{r\},\emptyset)dr + \\
  &z^2\int_0^t\int_0^{r_2}u_t(\{s\},\emptyset,\{r_1,r_2\},\emptyset)dr_1dr_2 = 
  \begin{pmatrix}z\exp(-\frac{t}{2})&2z^2\overline{\kappa}_f\exp(-\frac{t}{2})- 2z^2\overline{\kappa}_f\\
  \kappa_f\exp(-\frac{s}{2}) & z \end{pmatrix}.  
  \end{split}\end{equation*}
Therefore we get, for all $A \in M_2$:
  \begin{equation*}\begin{split}
  &\JC_f(A) = \lim_{t\downarrow 0} \frac{\int_0^t\Ad\big[U_t\pi(z)\ten\pi(0)(\{s\},\emptyset)\big](A)ds}
       {t\exp(-t|z|^2)} =                                                      
  \Ad\Bigg[\begin{pmatrix}z & 0 \\ \kappa_f & z\end{pmatrix}\Bigg](A) =                             \\
  &\Ad[zI_2+V_f](A).
  \end{split}\end{equation*}

Let us now turn to the calculation of $\JC_s$. We find for all $A \in M_2$:
  \begin{equation*}
  \JC_s(A) = \lim_{t\downarrow 0} \frac{1}{t} \EC^t\big[(\{\emptyset\},\Omega_1[-t,0))\big](A)
  = \lim_{t\downarrow 0} \frac{\int_0^t\Ad\big[U_t\pi(z)\ten\pi(0)(\emptyset,\{s\})\big](A)ds}
       {t\exp(-t|z|^2)}. 
  \end{equation*}
Taking a closer look at $U_t\pi(z)\ten\pi(0)(\emptyset,\{s\})$, applying Theorem \ref{Maassen}:
  \begin{equation*}\begin{split}
  &U_t\pi(z)\ten\pi(0)(\emptyset,\{s\}) =
  \int_{\Omega}u_t(\emptyset,\{s\},\tau,\emptyset)z^{|\tau|}d\tau =    
   u_t(\emptyset,\{s\},\emptyset,\emptyset)\ +                                                \\
  &z\int_0^t u_t(\emptyset,\{s\},\{r\},\emptyset)dr + 
  z^2\int_0^t \int_0^{r_2}u_t(\emptyset,\{s\},\{r_1,r_2\},\emptyset)dr_1dr_2 =
  \begin{pmatrix}0 &0 \\ \kappa_s\exp(-\frac{s}{2}) & 0 \end{pmatrix}. 
  \end{split}\end{equation*}
Therefore we get, for all $A \in M_2$:
  \begin{equation*}\begin{split}
  &\JC_s(A) = \lim_{t\downarrow 0} \frac{\int_0^t\Ad\big[U_t\pi(z)\ten\pi(0)(\emptyset,\{s\})\big](A)ds}
       {t\exp(-t|z|^2)} =                                                      
  \Ad\Bigg[\begin{pmatrix}0 & 0 \\ \kappa_s & 0 \end{pmatrix}\Bigg](A) =                            \\
  & = \Ad[V_s](A).    
  \end{split}\end{equation*}
Since we are driving the atom with a laser now, the time evolution when we do not observe
the side channel nor the forward channel is now given by $T^z_t := \EC^t\big[\Omega[-t,0),\Omega[-t,0)\big]$
and no longer by $T_t$. We will now derive the Master equation for this new time evolution.
For this we need the Dyson series: let $L_0$ and $J$ be maps from $M_2 \to M_2$, then for all $t \ge 0$:
\begin{equation*}
  \exp\big(t(L_0 +J)\big) = \int_{\Omega[-t,0)} \exp\big((\omega_1+t)L_0\big)J 
  \exp\big((\omega_2-\omega_1)L_0\big)J\ldots J\exp(-\omega_k L_0)d\omega,
\end{equation*}
where we have written $\omega$ as $\{\omega_1,\ldots, \omega_k\}$ 
with $-t \le \omega_1 < \ldots < \omega_k\le 0$.                                        \\
Now remember that $\{Y_t\}_{t\ge 0}$ is a semigroup so we can write $Y_t = \exp(tL_0)$.
Then, using Theorem \ref{deco} and twice the Dyson series, we see that:
  \begin{equation*}\begin{split}
  &T_t^z = \EC^t\big[\Omega[-t,0),\Omega[-t,0)\big] = 
  \int_{\Omega[-t,0)\times\Omega[-t,0)} W_{Y, \JC_f, \JC_s}(\omega_f,\omega_s)d\omega_f d\omega_s = \\
  &\exp\big(t(L_0+\JC_f +\JC_s)\big).
  \end{split}\end{equation*}
This means we get the following Master equation:
  \begin{equation}\label{Mastereq}
  \frac{d}{dt}T^z_t = L_0 + \JC_f + \JC_s = - \frac{1}{2}\{V^*V, \,\cdot\,\} + 
  [zV^*-\overline{z}V, \,\cdot\,] +V^* \,\cdot\,V,
  \end{equation}
which is exactly the Master equation for resonance fluorescence (see \cite{Car}) if we take
$z= -i \frac{\Omega}{2}$ with $\Omega$, the \emph{Rabi frequency}, real.                      \\
In the quantum optics literature (see for instance \cite{Car}), usually there is no photon counting
measurement done in the forward channel, i.e.\ $E_f = \Omega[-t,0)$. From here on we will do the same,
we define for all $t\ge0$ and $E_s \in \Sigma_t:\ \MC^t[E_s] := \EC^t\big[\Omega[-t,0), E_s\big]$.
In the following we will also suppress the index $s$ on $E_s$. Using the Dyson series and Theorem
\ref{deco} we find, for all $t\ge0$ and $E \in \Sigma_t$:
  \begin{equation}\label{Davside}
  \MC^t[E] = \int_E W_{Z, \JC_s}(\omega)d\mu(\omega),
  \end{equation} 
where the time evolution in between side-channel-jumps $Z_t$ is given by 
$Z_t = \exp\big(t(L_0+ \JC_f)\big)$ and $W_{Z, \JC_s}$ is defined in the obvious way analogous
to Definition \ref{W}. Note that we have found exactly the same jump operator and time evolution in
between jumps as in the usual quantum optics literature, see for instance \cite{Car}, \cite{Car3}, i.e.\
we have succeeded in constructing the Davies process of resonance fluorescence with quantum 
stochastic calculus.                                                                             \\

\section{A renewal process}

We will now look briefly at some features of the process $\MC^t$ we obtained. It is easily seen
from the fact that $(\JC_s)^2=0$ (i.e.\ $g_2(0) = 0$) that the photons in the side channel arrive
\emph{anti-bunched}: the probability to see two photons immediately after each other is $0$.
We will now show that the photon counting process in the side channel is a so-called
\emph{renewal process}.                                                              \\
We denote $\Sigma^t := \Sigma[0,t)$ and, via a shift, we let events $E$ in $\Sigma^t$ correspond to events
$E-t$ in the output sigma field $\Sigma_t$. This means that an element $\omega = \{\omega_1, \ldots, \omega_k\}$ 
in $E \in \Sigma^t$ with $0 \le \omega_1 < \ldots < \omega_k < t$ should be interpreted as 
seeing the first photon appear in the side channel at time $\omega_1$, the second at time $\omega_2$
up to the $k$'th photon at time $\omega_k$.                                                       \\
Given that we start the photon counting measurement in the initial state $\rho$, we define on the 
sigma fields $\Sigma^t$ ($t \ge 0$) probability measures in the natural way:
for $E \in \Sigma^t: \BB{P}^t_\rho[E] := \mbox{Tr}\big(\rho \MC^t[E-t](I)\big)$. The family of sigma fields
$\{\Sigma^t\}_{t\ge0}$ generates a sigma-field $\Sigma^{\infty}$ of $\Omega[0,\infty)$.
Using that $T_s^z(I) = I$, see equation \eqref{Mastereq}, we find for all $E \in \Sigma^t$:
  \begin{equation*}\begin{split}
  &\BB{P}^{t+s}_{\rho}[E] =
  \mbox{Tr}\Big(\rho \MC^{t+s}\Big[\big(E \tilde{\cup} \Omega[t,t+s)\big)-(t+s)\Big](I)\Big) =      \\
  &\mbox{Tr}\Big(\rho \MC^{t+s}\big[E-(t+s) \tilde{\cup} \Omega[-s,0)\big](I)\Big) =
  \mbox{Tr}\Big(\rho\MC^t[E-t]\MC^s\big[\Omega[-s, 0)\big](I)\Big) =                                        \\
  &\mbox{Tr}\big(\rho \MC^t[E-t]T_t^z(I)\big) = 
         \mbox{Tr}\big(\rho \MC^t[E-t](I)\big) = \BB{P}^{t}_{\rho}[E],
  \end{split}\end{equation*} 
so $\BB{P}^{t+s}_{\rho}[E]$ does not depend on $s$. Therefore the family 
$\{\BB{P}^t_{\rho}\}_{t\ge 0}$ on the sigma-fields $\{\Sigma^t\}_{t\ge 0}$
is consistent, hence by Kolmogorov's extension theorem it extends to a 
single probability measure $\BB{P}_{\rho}$ on $\Sigma^\infty$.                                        \\
In the following, when we write $\omega \in \Omega[0,\infty)$ as $\{\omega_{1}, \omega_2, \ldots \}$, we
imply that $0 \le \omega_1 < \omega_2 < \ldots$. For $j =1,2,\ldots$ we define random variables:
  \begin{equation*}
  X_j:\ \Omega[0,\infty) \to \overline{\BB{R}}^+:\ \omega = \{\omega_1, \omega_2,\ldots\}
  \mapsto
  \left\{ \begin{array}{ll}
  \omega_j-\omega_{j-1} & \mbox{\ if\ \ \ }  |\omega| \ge j \\
  \infty & \mbox{\ else\ } \end{array}\right. ,
  \end{equation*}
where we take $\omega_0$ to be $0$. These random variables give the time elapsed between 
the $(j-1)$th and $j$th detection of a photon. To prove that the counting process is a 
\emph{(modified) renewal process} we have to show that for $i= 1,2,\ldots$
the random variables $X_i$ are independent and for $i = 2,3,\ldots$ they are identically distributed. 
This means we have to show that for $i = 2,3,\ldots$ the
distribution functions $F_{X_i}(x) := \BB{P}_\rho[X_i \le x]$ are equal and for $i,j = 1,2,\ldots$
the joint probability distribution function $F_{X_i, X_j}(x,y) := \BB{P}_\rho[X_i \le x \wedge X_j\le y]$
factorizes: $F_{X_i, X_j}(x,y) = F_{X_i}(x)F_{X_j}(y)$.                                                \\
Let us first introduce some convenient notation. Note that, using equation \eqref{Davside}, we 
have for all $E \in \Sigma^t$:
  \begin{equation*}\begin{split}
  &\BB{P}_\rho[E] = \BB{P}^t_\rho[E] = \mbox{Tr}\Big(\rho\int_{E-t}W_{Z,\JC_s}(\omega)d\mu(\omega)(I)\Big) =  \\
  &\mbox{Tr}\Big(\rho\int_{E}Z_{\omega_1}\JC_s Z_{\omega_2-\omega_1}\JC_s \ldots 
  \JC_s Z_{t-\omega_k}(I)d\mu(\omega)\Big).
  \end{split}\end{equation*}
We will denote: $x_1 := \omega_1, x_2 := \omega_2-\omega_1, \ldots, x_{k+1} := t- \omega_k$, then we can 
write:
  \begin{equation*}
  \BB{P}_\rho[E] = \int_E \mbox{Tr}\Big(\rho Z_{x_1}\JC_s Z_{x_2}\JC_s \ldots 
  \JC_s Z_{x_{k+1}}(I)\Big)d\mu(\omega). 
  \end{equation*}
Let $P$ denote the matrix $\begin{pmatrix}1 & 0 \\ 0 & 0\end{pmatrix}$, then we have: 
  \begin{equation*}\begin{split}
  &\JC_s Z_{x_{k+1}}(I) = \begin{pmatrix}|\kappa_s|^2\big(Z_{x_{k+1}}(I)\big)_{22} & 0 \\
  0 & 0 \end{pmatrix} = |\kappa_s|^2\big(Z_{x_{k+1}}(I)\big)_{22} P,                          \\
  &\JC_s Z_{x_{k}}(P) = \begin{pmatrix}|\kappa_s|^2\big(Z_{x_{k}}(P)\big)_{22} & 0 \\
  0 & 0 \end{pmatrix} = |\kappa_s|^2\big(Z_{x_{k}}(P)\big)_{22} P,  \ldots,                   \\
  &\JC_s Z_{x_{2}}(P) = \begin{pmatrix}|\kappa_s|^2\big(Z_{x_{2}}(P)\big)_{22} & 0 \\
  0 & 0 \end{pmatrix} = |\kappa_s|^2\big(Z_{x_{2}}(P)\big)_{22} P.
  \end{split}\end{equation*}
Therefore, if we define $z(x) := |\kappa_s|^2\big(Z_x(P)\big)_{22}$,  
$z_{last}(x) := |\kappa_s|^2\big(Z_x(I)\big)_{22}$ and $z_{first}(x) := \mbox{Tr}\big(\rho Z_x (P)\big)$,
we can write (see also \cite{Car3}): 
  \begin{equation}\label{formula}
  \BB{P}_\rho[E] = \int_E
   z_{first}(x_1)\Big(\prod_{l=2}^k z(x_l)\Big)z_{last}(x_{k+1})d\mu(\omega). 
  \end{equation}
We would like to stress that this formula is only valid for events $E \in \Sigma^t$ and not 
for all events in $\Sigma^\infty$.                                                                \\
For $t \ge 0$ we introduce the following random variables:
  \begin{equation*}
  N_t:\ \Omega[0,\infty) \to \BB{N}:\ \omega \mapsto |\omega \cap [0,t] |,
  \end{equation*}  
counting the number of photons appearing in the side channel up to time $t$. Since, for strictly positive
driving field strengths, i.e.\ $|z| >0$,
the eigenvalues of the generator $L_0 + \JC_f$ of the semigroup $Z_t$ all have strictly negative real parts, 
we have $\lim_{t\to\infty}Z_t = 0$. Using this, formula \eqref{formula} and the fact that 
the event $[N_t =0]$ is an element of $\Sigma^t$, we obtain:
  \begin{equation*}
  \lim_{t\to \infty} \BB{P}_\rho[N_t = 0] = \lim_{t \to \infty} z_{first}(t) = 0. 
  \end{equation*}
Now suppose we have that $\lim_{t\to \infty} \BB{P}_\rho[N_t \le n] = 0$ for a certain $n \in \BB{N}$. 
For $s \le t$ we use: $\BB{P}_\rho[N_t \le n+1] = 
\BB{P}_\rho[N_t \le n+1| N_s \le n ]\BB{P}_\rho[N_s \le n] +
\BB{P}_\rho[N_t \le n+1| N_s > n]\BB{P}_\rho[N_s > n]$. Therefore we have:
  \begin{equation*}\begin{split}
  &\lim_{t\to \infty} \BB{P}_\rho[N_t \le n +1] = \lim_{s\to \infty}\lim_{t\to \infty} 
  \BB{P}_\rho[N_t \le n +1] =                                                                 \\
  &\lim_{s\to \infty}\lim_{t\to \infty}\Big(
  \BB{P}_\rho[N_t \le n+1| N_s \le n ]\BB{P}_\rho[N_s \le n] + 
  \BB{P}_\rho[N_t \le n+1| N_s > n ]\BB{P}_\rho[N_s > n]\Big) =                               \\
  &\lim_{s\to \infty}\lim_{t\to \infty}\BB{P}^t_\rho[N_t \le n+1| N_s > n ] = 
  \lim_{s\to \infty}\lim_{t\to \infty}z_{last}(t-s) = 0.
  \end{split}\end{equation*}
Now using induction, we get for $n \in \BB{N}$:
  \begin{equation*}
  \lim_{t \to \infty} \BB{P}_\rho [N_t \le n] = 0.
  \end{equation*}
We are now ready to calculate the distribution functions $F_{X_i}$ and $F_{X_i, X_j}$. The 
problem is that for instance the event $[X_i \le x] \in \Sigma^\infty$ is not an element of 
$\Sigma^t$ for a $t \in \BB{R}$. We solve this by conditioning on the event $[N_t \ge i]$
and taking the limit for $t$ to infinity:
  \begin{equation*}\begin{split}
  &F_{X_i}(x) = \BB{P}_\rho [X_i \le x] =                                                          \\
  &\lim_{t \to \infty} \Big(\BB{P}_\rho [X_i \le x| N_t \ge i] \BB{P}_\rho[N_t \ge i]+ 
   \BB{P}_\rho [X_i \le x| N_t < i]\BB{P}_\rho[N_t < i]\Big) =                                         \\
  &\lim_{t \to \infty}\BB{P}^t_\rho [X_i \le x \wedge N_t \ge i].
  \end{split}\end{equation*}
Now we use again formula \eqref{formula} to obtain for $i \ge 2$:
  \begin{equation*}\begin{split} 
  &F_{X_i}(x) =  \lim_{t \to \infty} \sum_{k=i}^\infty \int_{\substack{\sum_{l=1}^{k+1} x_l = t \\ x_i \le x}}
  z_{first}(x_1)\Big(\prod_{l=2}^k z(x_l)\Big)z_{last}(x_{k+1})dx_1\ldots dx_{k+1} =                      \\
  &\lim_{t \to \infty} \int_0^x z(x_i)\Bigg(\sum_{k=i}^\infty 
   \int_{\sum_{l \neq i} x_l = t-x_i}
  z_{first}(x_1)dx_1\Big(\prod_{\substack{l =2\\ l\neq i}}^k z(x_l)dx_l\Big)z_{last}(x_{k+1})dx_{k+1}\ 
  \Bigg)dx_i =                                                                                             \\
  &\lim_{t \to \infty} \int_0^x z(x_i) \BB{P}^{t-x_i}_\rho\big[N_{t-x_i} \ge i-1\big] dx_i.
  \end{split}\end{equation*}
Then we use dominated convergence to interchange the limit and the integral to obtain:
  \begin{equation*}
  F_{X_i}(x) = \int_0^x z(x')dx'.
  \end{equation*}
When $i=1$ we can repeat the whole calculation to find the same result when for $z$ we substitute
$z_{first}$. It is now obvious that for $i = 2,3,\ldots$ the random variables $X_i$ are identically
distributed.                                                                                        \\
In a similar fashion, only extracting two integrals now, we find that for $i,j = 2,3, \ldots:\
F_{X_i,X_j}(x,y) = \int_0^x\int_0^y z(x')z(y') dx'dy'$. If $i$ or $j$ is $1$ we again have to 
substitute $z_{first}$ for $z$. It is now obvious that the random variables $X_i$ and $X_j$ are
independent. We conclude that the family of random variables $\{X_i\}_{i=1,2,\ldots}$ is a 
(modified) renewal process.                                                                       \\

\textbf{Acknowledgement:}
L.B.\ would like to thank Howard Carmichael for hospitality and discussion on 
the topic of this article while visiting his group in Oregon.

}
\end{document}